\documentclass[twocolumn,amsmath,amssymb,prl]{revtex4}
\usepackage[dvips]{graphicx}
\usepackage{dcolumn}
\usepackage{bm,soul}
\newcounter{one}
\usepackage{braket}
\usepackage[usenames]{color}
\usepackage{float}

\newcommand{\tol}[1]{#1.}

\definecolor{nblue}{rgb}{0.3,0.3,1.0}
\definecolor{ngreen}{rgb}{0.2,0.7,0.2}
\definecolor{nred}{rgb}{0.9,0.1,0}
\definecolor{norange}{rgb}{0.8,0.5,0}

\newcommand{\beq}{\begin{equation}}
\newcommand{\eeq}{\end{equation}}
\newcommand{\bqa}{\begin{eqnarray}}
\newcommand{\eqa}{\end{eqnarray}}

\newcommand{\rt}[1]{\sqrt{#1}\,}
\newcommand{\erf}[1]{Eq.~(\ref{#1})}

\newcommand{\erfand}[2]{Eqs.~(\ref{#1}) and (\ref{#2})}

\newcommand{\ip}[2]{\left\langle{#1}\right|\left.\!\!{#2}\right\rangle}

\newcommand{\sch}{Schr\"odinger}

\newcommand{\cu}[1]{\left\{ {#1} \right\}}

\begin{document}

\title{Experimental Proof of Nonlocal Wavefunction Collapse for a Single Particle Using Homodyne Measurement}

\author{Maria Fuwa${}^{1}$}
\author{Shuntaro Takeda${}^{1}$}
\author{Marcin Zwierz${}^{2,3}$}
\author{Howard M. Wiseman${}^{3}$}\email{h.wiseman@griffith.edu.au}
\author{Akira Furusawa${}^{1}$}\email{akiraf@ap.t.u-tokyo.ac.jp}
\affiliation{${}^1$Department of Applied Physics, School of Engineering, The University of Tokyo,\\ 7-3-1 Hongo, Bunkyo-ku, Tokyo 113-8656, Japan\\  ${}^2$Faculty of Physics, University of Warsaw, Pasteura 5, 02-093 Warsaw, Poland\\ ${}^3$Centre for Quantum Computation and Communication Technology (Australian Research Council), Centre for Quantum Dynamics, Griffith University, Brisbane, QLD 4111, Australia}

\date{\today}

\begin{abstract}	
A single quantum particle can be described by a wavefunction that spreads over arbitrarily large distances, but it is never detected in two (or more) places. This strange phenomenon is explained in quantum theory by what Einstein repudiated as ``spooky action at a distance'': the instantaneous nonlocal collapse of the wavefunction to wherever the particle is detected. We demonstrate this single-particle spooky action, for the first time with no efficiency loophole, by splitting a single photon between two laboratories and experimentally testing if the choice of
 measurement in one lab really causes a change in the local quantum state in the other lab.
To this end, we use homodyne measurements with six different measurement settings and quantitatively verify Einstein's spooky action by violating an Einstein-Podolsky-Rosen-steering inequality by $0.042 \pm 0.006$.
Our experiment also verifies the entanglement of the split single photon even when one side is untrusted.
\end{abstract}

\maketitle

Einstein never accepted orthodox quantum mechanics because he did not believe that its nonlocal collapse of the wavefunction could be real. When he first made this argument in 1927~\cite{Bacci09}, he considered just a single particle.
The particle's wavefunction was diffracted through a tiny hole so that it ``dispersed'' over a large hemispherical area prior to encountering a screen of that shape covered in photographic film.  Since the film only ever registers the particle at one point on the screen, orthodox quantum mechanics must postulate  a ``peculiar mechanism of action at a distance, which prevents the wave \ldots\ from producing an action in two places on the screen''~\cite{Bacci09}. That is, according to the theory, the detection at one point must instantaneously collapse the wavefunction to nothing at all other points.

Here, for the first time, we demonstrate Einstein's ``spooky action at a distance'' \cite{Born49}
using a single particle (a photon),  as in his original conception.
We simplify Einstein's 1927 {\em gedankenexperiment} so
that the single photon is split into just two
wavepackets, one sent to a lab supervised by Alice and the other to a distant lab supervised by Bob.
But there is a key difference, which enables us to demonstrate the nonlocal collapse experimentally:
rather than simply detecting the presence or absence of the photon, we use
homodyne measurements.
 This gives Alice the power to make different measurements,
 and enables Bob to test (using tomography) whether Alice's measurement choice
 affects the way his conditioned state collapses,
 without having to trust anything outside his own laboratory.

The key role of measurement choice by Alice in demonstrating ``spooky action at a distance''
 was introduced in the famous Einstein-Podolsky-Rosen (EPR) paper~\cite{PhysRev47-777} of 1935.
In its most general form~\cite{PRL98-140402}, this phenomenon has been called
 EPR-steering~\cite{PRA80-032112}, to acknowledge the contribution and terminology of
 \sch~\cite{ProcCambridlePhilosSoc31-553}, who talked of Alice ``steering'' the state of Bob's quantum system.
 From a quantum information perspective, EPR-steering is equivalent to the task of entanglement
 verification when Bob (and his detectors) can be trusted but Alice (or her detectors) cannot \cite{PRL98-140402}.
This is strictly harder than verifying entanglement with both parties trusted~\cite{Terhal00}, but strictly  easier than violating a Bell inequality~\cite{Bell64}, where neither party is trusted~\cite{Terhal00}.

To demonstrate EPR-steering quantitatively it is necessary and sufficient to violate an EPR-steering inequality involving Alice's and Bob's results~\cite{PRA80-032112}.
 Such a violation has been shown to be necessary for one-sided device-independent quantum key distribution as well~\cite{PRA85-010301}.
Because Alice is not trusted in EPR-steering, a rigorous experiment cannot use postselection on Alice's side~\cite{PRA80-032112, PRL98-140402, NatCom3-625, NewJPhys14-053030, PRX2-031003, EvansPRA2013}. 
Previous experimental tests of nonlocal quantum state collapse over macroscopic distances,
 without postselection on Alice's side, have involved the distribution of entangled states of multiple particles~\cite{RevModPhys81-1727, arXiv12086222, NatPhys6-845, NatCom3-625, NewJPhys14-053030,  PRX2-031003, NewJPhys14-113020, PRL68-3663, PRA90-043601, NatPhoton6-596, PRL106-130402, PRA87-022104, PRL110-130407, JOptB5-S467, PRA83-052329, PRL92-210403}.
Experiments demonstrating Bell-nonlocality (violating a Bell inequality) for a single photon have involved postselection on both sides, opening the efficiency loophole~\cite{PRL92-193601, PRL92-180401}; these works would  otherwise have demonstrated EPR-steering of a single photon as well. A recent experimental test  of entanglement for a single photon via an entanglement witness has no efficiency loophole~\cite{PRL110-130401} but it demonstrates a weaker form on nonlocality than EPR-steering~\cite{PRA80-032112, PRL98-140402}. 
Our experiment is the first rigorous demonstration of this ``spooky action at a distance" for a single particle without opening the efficiency loophole.

While the nonlocal properties of a single particle have spurred much theoretical debate~\cite{PRL66-252, PRL73-2279, PRA63-012305, PRA64-042106, PRL91-097902, PRA71-032339, PRL99-180404, PRA84-012110, JPhysA45-215308, PRA88-012111} and many fundamental experiments~\cite{PRL88-070402, PRA66-024309, PRL92-047903, arxiv12041712, PRL92-193601, PRL92-180401, PRL110-130401, PRL104-180504},
it is also recognized that
a single photon split between two spatially distant modes is a very flexible entanglement resource for quantum information tasks: they have been teleported~\cite{PRL88-070402}, swapped~\cite{PRA66-024309} and purified with linear optics~\cite{PRL104-180504}.
Spatial-mode entanglement~\cite{PRL91-097902} more generally has a broad potential ranging from long distance quantum communication~\cite{RevModPhys83-33, Nature414-413}, quantum computation~\cite{Nature396-46, PRL88-070402, PRA66-024309, PRL104-180504}, to the simulation of quantum many-body systems in tabletop implementations~\cite{NatPhys2-803}.
Our work is the first one-sided device-independent verification of this type of entanglement.

\section{Results}

\begin{figure}[!b]
\begin{center}
\includegraphics[bb=127 128 497 357, width= \linewidth ]{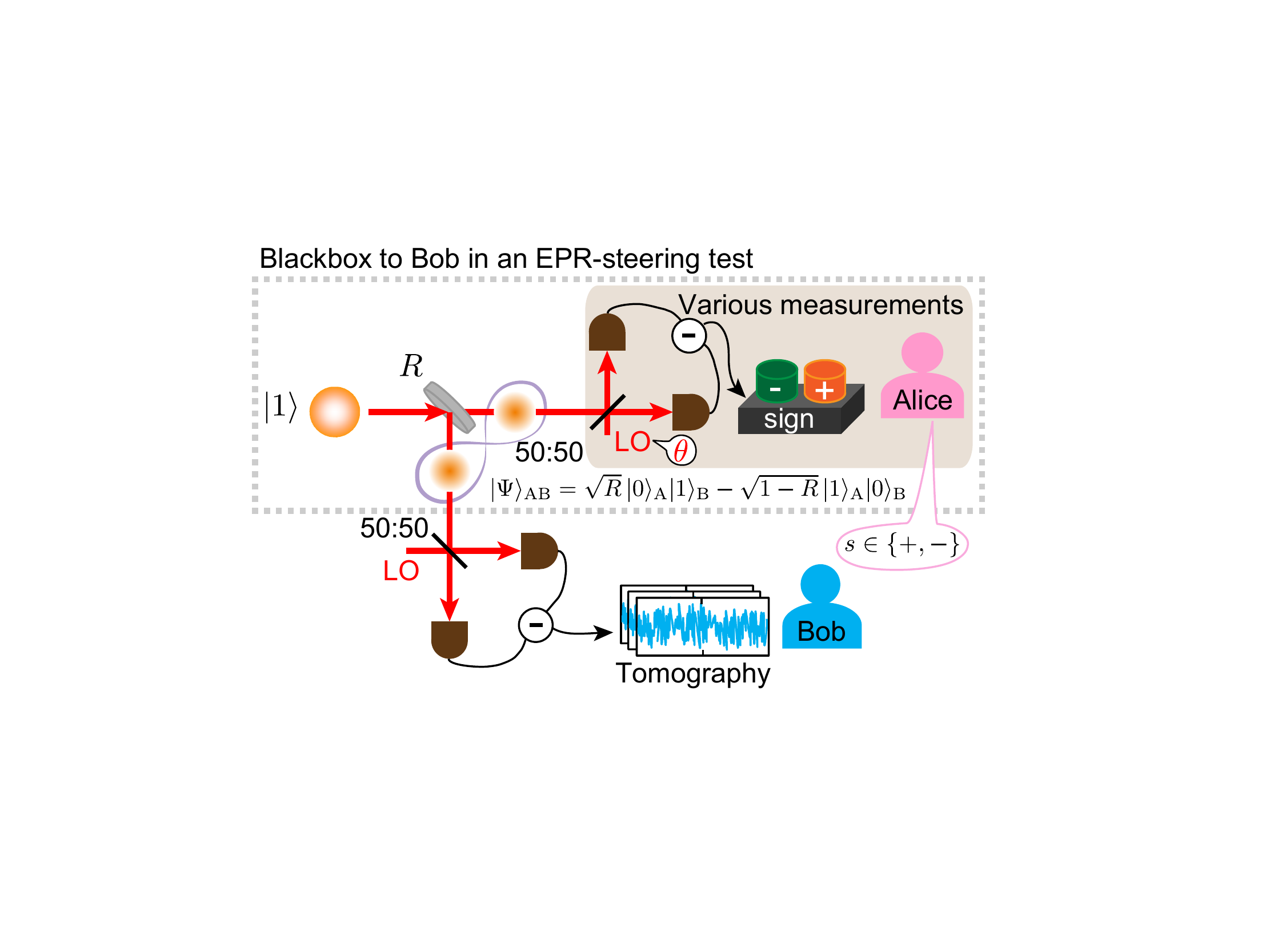}
\end{center}
\caption{Our simplified version of Einstein's original \textit{gedankenexperiment}.  A single photon is  incident on a beam splitter of reflectivity $R$ and then subjected to homodyne measurements at two spatially separated locations. Alice is trying to convince Bob that she can steer his portion of the single photon to different types of local quantum states
by using different values of the local oscillator (LO) phase $\theta$ on her side.}
\label{fig-overall_schematic}
\end{figure}

\begin{figure*}[!t]
\begin{center}
\includegraphics[bb=134 119 1057 717, width=\linewidth ]{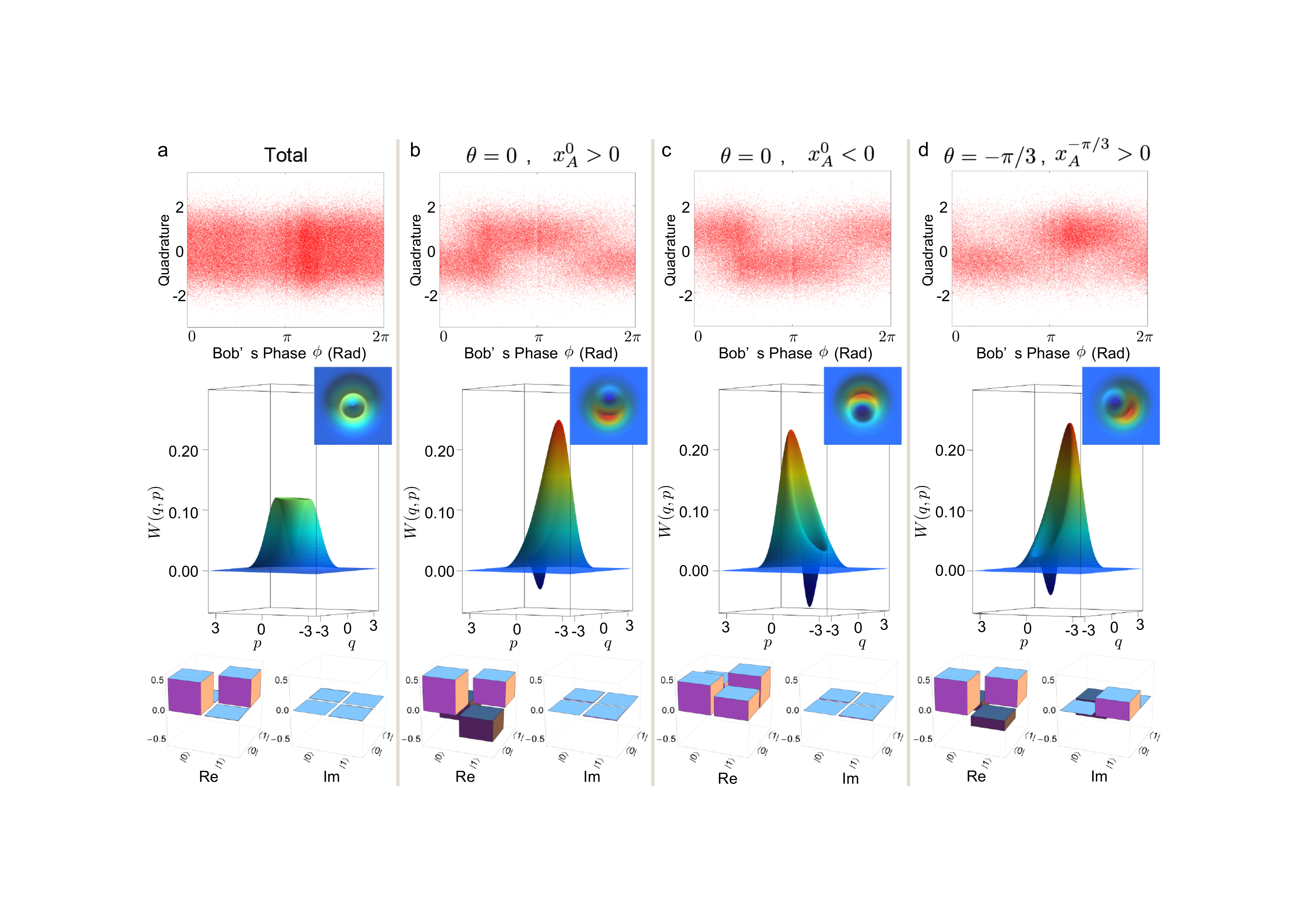}
\end{center}
\vspace{-4mm}
\caption{Bob's unconditioned and conditioned quantum states for  $R=0.50$. (a) Bob's unconditioned data.
(b) Bob's data conditioned on $\theta = 0$, positive quadrature sign $s$ ($x^{0}_{\mathrm{A}} > 0$). (c) $\theta = 0$, negative quadrature sign $s$ ($x^{0}_{\mathrm{A}} < 0$). (d) $\theta = -\pi/3$, positive quadrature sign $s$
($x^{-\pi/3}_{\mathrm{A}} > 0$).
(Top) Bob's marginal distributions (the variation in the number of data points as a function of Bob's LO phase $\phi$ is due to unevenness in Bob's phase scanning, and does not affect the reconstructions).
(Middle) Reconstruction of Wigner functions (top view in the inset); here $\hbar = 1$ and $(q,p)$ label
canonically conjugate quadratures.
(Bottom) Reconstruction of truncated density matrices defined on the $\{ \ket{0}, \ket{1}\} $ subspace; negative elements in darker shade.}
\label{fig:state_results}
\end{figure*}

\textbf{Our gedankenexperiment} 
First we explain in detail our simplified version (Fig.~\ref{fig-overall_schematic}) of Einstein's original single-particle \textit{gedankenexperiment} described above, formalized as the task of entanglement verification with only one trusted party, as proposed in Ref.~\cite{PRA84-012110}. That is, assuming only that Bob can reliably probe the quantum state at his location, he can experimentally prove that the choice of measurement by the distant Alice affects his quantum state. This is exactly the ``spooky action at a distance'' that Einstein found objectionable~\cite{born-einstein}.

We start with a pure single photon $\ket{1}$ incident on a beam splitter of reflectivity $R$. As a result, the state of the single photon becomes spread out between two spatially separated modes A and B:
\begin{equation}
\ket{\Psi}_{\mathrm{AB}} = \sqrt{R} \, \ket{0}_{\mathrm{A}} \ket{1}_{\mathrm{B}} - \sqrt{1-R} \, \ket{1}_{\mathrm{A}} \ket{0}_{\mathrm{B}}.
\label{pure_split_photon}
\end{equation}
The transmitted mode is sent to Alice, and the reflected one to Bob. 
We allow Alice and Bob to use homodyne detection. This allows Bob to do quantum tomography on
his state~\cite{Vogel89, Lvovsky09}, and gives Alice the power to do different types of measurement (which is necessary for
an EPR-steering test) by controlling her local oscillator (LO) phase $\theta$.

If Alice were simply to detect the presence or absence of a photon, then Bob's measurement of
the same observable will be anticorrelated with Alice's, as in Ref.~\cite{arxiv12041712}.
But this does not prove that Alice's measurement affected Bob's local state, because such perfect
anticorrelations would also arise from a classical mixture of $\ket{0}_{\mathrm{A}} \ket{1}_{\mathrm{B}}$
and $\ket{1}_{\mathrm{A}} \ket{0}_{\mathrm{B}}$, in which Bob's measurement simply reveals a pre-existing
local state for him, $\ket{1}_{\mathrm{B}}$ or $\ket{0}_{\mathrm{B}}$. To demonstrate nonlocal quantum state collapse,
measurement choice by Alice is essential~\cite{PRL98-140402}. 

Following Alice's homodyne measurement of the $\theta$-quadrature $X^\theta_{\mathrm{A}}$,
yielding result $x^\theta_{\mathrm{A}} \in \mathbb{R}$,
Bob's local state is collapsed to 
\begin{align} 
\ip{x^\theta_{\mathrm{A}}}{\Psi}_{\mathrm{AB}} & \propto \sqrt{R} \,\ket{1}_{\mathrm{B}} - \sqrt{1-R} \, e^{-i \theta} \rt{2} x^\theta_{\mathrm{A}} \ket{0}_{\mathrm{B}},
\label{bob_conditional_qubit_approx}
\end{align}
where the proportionality factor is $c(x^{\theta}_{\mathrm{A}}) = \exp[-(x^{\theta}_{\mathrm{A}})^{2}/2]/\sqrt[4]{\pi}$.  
Thus by changing her LO phase $\theta$, Alice controls the relative phase of the vacuum and one photon component of Bob's
conditioned state (modulo $\pi$, depending on the sign of the $x^\theta_{\mathrm{A}}$ she obtains). Because of this,
it is convenient for Alice to coarse-grain her result to $s(x^\theta_{\mathrm{A}})={\rm sign}(x^\theta_{\mathrm{A}}) \in \cu{+,-}$.
It is possible that a more sensitive EPR-steering inequality could be obtained that makes use of a finer-grained binning of Alice's results, but two bins are sufficient for our experiment. 

 Independently of Alice's measurement, Bob performs full quantum state tomography using
homodyne detection on his portion of the single photon. This enables him to reconstruct his state, for each value of Alice's LO setting $\theta$, and coarse-grained result $s$. Because of the coarse-graining,
even under the idealisation of the pure state as in \erf{bob_conditional_qubit_approx},
Bob's (normalized) conditioned state will be mixed:
\begin{eqnarray}\label{idealizedconditionedstate}
\hat{\rho}^{\theta}_{s} &=& R\ket{1}\bra{1} + (1-R)\ket{0}\bra{0}
 \nonumber \\
&-& s \sqrt{R(1-R)2/\pi} (e^{i \theta}  \ket{1}\bra{0} + e^{-i \theta} \ket{0}\bra{1}).
\end{eqnarray}
The idealized theoretical prediction for the unconditioned quantum state is
\begin{equation}\label{mixture}
\hat{\rho} = \sum_{s} P(s|\theta) \hat{\rho}^{\theta}_{s} = R\ket{1}\bra{1} + (1-R)\ket{0}\bra{0},
\end{equation}
where $P(s|\theta) = 0.5$ is the relative frequency for Alice to report sign $s$ given setting $\theta$.

We once again emphasize the intrinsic lack of trust that Bob has with respect to anything that happens in Alice's lab. Neither her honesty nor the efficiency or accuracy of her measurement devices is assumed in an EPR-steering test. On the other hand, Bob does trust his own measurement devices. From the experimental point of view this means that his photoreceivers do not have to be efficient,  and that he can post-select on finding his system in a particular subspace.
In particular, for our experiment (where there are small two-photon terms),
 he can restrict his reconstructed state $\hat{\rho}_s^\theta$ to the qubit subspace spanned by
$\cu{\ket{0}_{\mathrm{B}},\ket{1}_{\mathrm{B}}}$.
Despite Bob's lack of trust in Alice, she can convince him that her choice of measurement setting, $\theta$, steers
his quantum state $\hat{\rho}_s^\theta$, proving that his system has no local quantum description. We now present data
showing this effect qualitatively, prior to our quantitative proof of EPR-steering for this single-photon system.

\textbf{Bob's Tomography Results}
In our experiment, the (heralded single-photon) input state to the beam splitter
comprises mostly a pure single-photon state $|1\rangle$, but it has some admixture of the vacuum state $|0\rangle$, and a (much smaller) admixture of the two-photon state $|2\rangle$.
(for more details on state preparation see the METHODS). Following the beam splitter and Alice's measurement, Bob reconstructs his conditioned quantum states by separately analyzing his homodyne data (taken by scanning his LO phase $\phi$ from $0$ to $2\pi$) for each value of Alice's LO phase $\theta$ and result $s$. The reconstructed density matrices $\hat{\rho}_s^\theta$ in the $\cu{\ket{0}_{\mathrm{B}},\ket{1}_{\mathrm{B}}}$ subspace and the corresponding Wigner functions give complementary ways to
to visualize how Bob's local quantum state can collapse in consequence of Alice's measurement.

The results of Bob's tomography, which take into account inefficiency in Bob's detection system by using the maximum likelihood method~\cite{Lvovsky09} during quantum state reconstruction, are presented in Fig.~\ref{fig:state_results},  for the case $R=0.50$.
There is good qualitative agreement between these results and the theoretical predictions for the ideal
case in \erfand{idealizedconditionedstate}{mixture}. In particular, there are four features to note.

First, Bob's unconditioned quantum state is a phase-independent statistical mixture of the vacuum and single-photon components [see Fig.~\ref{fig:state_results}(a)], as in \erf{mixture}. The Wigner functions are rotationally invariant, and the off-diagonal terms in $\hat\rho$ are zero. The vacuum component $p^{\mathrm{u}}_{0} = 0.55$ 
is slightly greater than the single-photon component $p^{\mathrm{u}}_{1} = 0.45$ 
due to the less than unit efficiency  $p_{1} = 0.857 \pm 0.008 < 1$ of single-photon generation (for more details see the METHODS). 

Second, Bob's conditioned quantum states are not phase-independent, but rather exhibit coherence between the
vacuum and single-photon components [see Fig.~\ref{fig:state_results}(b)--(d)] as predicted by \erf{idealizedconditionedstate}.
The  Wigner functions are not rotationally invariant (and have a mean field: $\langle q+ip\rangle \neq 0$),
 and the off-diagonal terms in $\hat\rho$ are non-zero.
Furthermore, the negative dips observed in the conditioned Wigner functions prove the strong nonclassical character of Bob's local quantum state~\cite{Leonhardt05}.

Third, depending on Alice's result $s \in \cu{+, -}$, Bob's local quantum state is collapsed into complementary states
(in the sense that they sum to the unconditioned state) [compare columns (b) and (c) in
Fig.~\ref{fig:state_results}.]
This effect is manifested most clearly by a relative $\pi$ rotation between the conditioned Wigner functions $W^{0}_{s} (q,p)$ and the opposite signs of the off-diagonal elements of $\hat{\rho}^{0}_{s}$, as expected from Eq.~(\ref{idealizedconditionedstate}).

Finally, Alice can steer Bob's possible conditioned states by her choice of measurement setting $\theta$,
as predicted. Comparing the results in columns (b) and (d) of Fig.~\ref{fig:state_results}, it is immediately clear that the conditioned Wigner function $W^{-\pi/3}_{+} (q,p)$ is phase shifted with respect to $W^{0}_{+} (q,p)$ by an angle $\theta = -\pi/3$. Moreover, we also notice the decrease in the value of the real off-diagonal elements and the emergence of the imaginary off-diagonal elements in the conditioned density matrix $\hat{\rho}^{-\pi/3}_{+}$ as compared to $\hat{\rho}^{0}_{+}$.
Naturally, the described EPR-steering effect can be demonstrated for all possible values of Alice's LO phase $\theta$.

The above results suggest that Bob's portion of the single photon cannot have a local quantum state prior to Alice defining her measurement setting $\theta$. However a proof that this is the case requires much more quantitative analysis, which we now present.

\begin{figure}[!t]
\begin{center}
\includegraphics[bb=115 426 458 633, width=\linewidth]{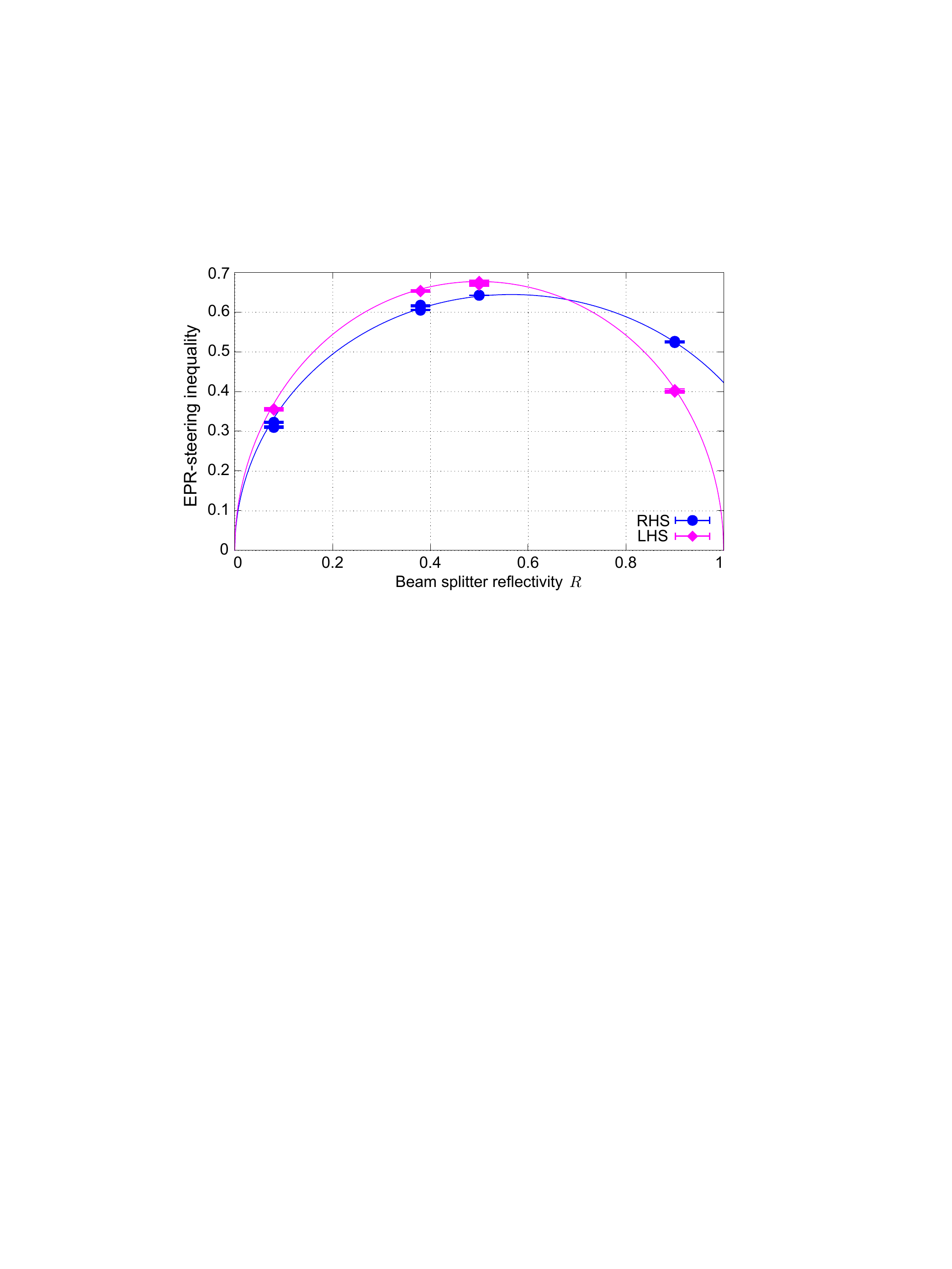}
\end{center}
\vspace{-4mm}
\caption{The left- and right-hand sides (LHS, RHS) of the EPR-steering inequality given in \erf{JWineq} plotted as functions of the beam splitter reflectivity $R$. The solid lines show the theoretically predicted values, which take into the account the experimental losses, the imperfections associated with the single-photon source, and the imperfections of Alice's homodyne measurements such as phase fluctuations. Four sets of data were taken for each $R$ to confirm reproducibility of the experimental results. To assess the tomographic errors, error bars were calculated using the bootstrap method, a widely used technique of resampling the data to estimate confidence intervals; for details see~\cite{bootstrap_Efron}.}
\label{fig:inequality_results}
\end{figure}

\textbf{The EPR-steering inequality}
From \erf{bob_conditional_qubit_approx}, Bob's portion of the single photon is a qubit (a quantum system spanned by $\ket{0}_{\mathrm{B}}$ and $\ket{1}_{\mathrm{B}}$).
In the experiment there are small terms with higher photon numbers, but, as explained above, Bob is allowed to restrict to
 the $\{\ket{0}_{\mathrm{B}}, \ket{1}_{\mathrm{B}}\}$  subspace. Here we consider a nonlinear EPR-steering inequality for the qubit subspace. It involves $n$ different measurement settings $\theta_{j}$ by Alice,
 and is given by~\cite{PRA84-012110}
\begin{equation} \label{JWineq}
\frac{1}{n} \sum_{{j}=1}^{n} \sum_s P(s|\theta_{j})  \, s \, {\rm Tr}[\hat{\sigma}^{\theta_{j}}_{\mathrm{B}} \hat{\rho}_s^{\theta_{j}}] \leq f(n) \sqrt{1 - \mathrm{Tr}\left[ \hat{\sigma}_{\mathrm{B}}^{\mathrm{z}} \hat{\rho} \right]^{2}}.
\end{equation}
Here
 $\hat{\sigma}^{\theta} \equiv e^{-i \theta}\ket{1}\bra{0} + e^{i \theta}  \ket{0} \bra{1}$
 and $f(n)$ is a monotonically decreasing positive function of the number of measurement settings
 defined in Eq. (4.15) of Ref.~\cite{PRA84-012110}, under the assumption that $\theta_{j} = \pi({j}/n)$.
The left-hand-side thus correlates Bob's tomographic reconstruction with Alice's announced result $s$,
but makes no assumptions about how Alice generates this result.
On the right-hand-side, $\hat{\rho}$ is Bob's unconditioned state, while $\hat{\sigma}_{\mathrm{B}}^{\mathrm{z}}=\ket{1}\bra{1}-\ket{0}\bra{0}$.

For the ideal case considered above, theory predicts a violation of the EPR-steering inequality (\ref{JWineq})
for $n\geq 2$ and any value of $R$ (apart from $0$ and $1$). [However, experimental imperfections associated with the single-photon source and the inefficiency of Alice's photoreceivers make it more difficult to obtain a violation. For details
of the theoretical predictions, see Supplementary section 2]
While the inequality is most easily violated for $n=\infty$ [for which $f(\infty)=2/\pi \approx 0.6366 $]
for our experiment it was sufficient to use $n=6$ [for which $f(6) = 0.6440$].
 The experimental results in Fig.~\ref{fig:inequality_results}
 well match the theoretical predictions calculated using independently measured experimental parameters; see Supplementary sections 3 and 4. The EPR-steering inequality is violated for $R=0.08, \, 0.38$, and $0.50$, but not for $R=0.90$; it is most violated at $R=0.38$ by $0.042\pm0.006$.

The violation of the EPR-steering inequality by seven standard-deviations
is a clear proof that Bob's quantum state cannot exist independently of Alice, but rather is collapsed by Alice's measurement.
We were able to rigorously demonstrate this for the first time for a single particle without opening the efficiency loophole by using the combination of multiple ($n=6$) measurement settings and highly efficient phase-sensitive homodyne measurements for Alice ($\eta_h = 0.96 \pm 0.01$), coupled with a high single-photon occupation probability ($p_{1} = 0.857 \pm 0.008 $).
Without the close-to-unity values of these parameters,
the nonlocal collapse of the single photon wavefunction could not have been detected,
as in the case of Ref.~\cite{PRL92-047903} (see Ref.~\cite{PRA84-012110} for a detailed discussion).

\section{Discussion}
We have demonstrated, both rigorously and in the easy visualized form of nonclassical Wigner functions, the nonlocality of a single particle using a modern and simplified version of Einstein's original \textit{gedankenexperiment}. That is, we demonstrated Einstein's ``spooky action at a distance'' in that
Bob's quantum state (of his half of a single photon) was provably dependent on Alice's choice of measurement
(on the other half), and could not have been pre-existing.
Quantitatively, we violated a multi-setting nonlinear EPR-steering inequality by several standard deviations ($0.042\pm0.006$).

This EPR-steering experiment is a form of entanglement verification, for a single-photon mode-entangled state,
which does not require Bob to trust Alice's devices, or her reported outcomes.
It was possible only because we used a high-fidelity single photon state and very high efficiency homodyne measurements,  to perform the steering measurements on Alice's side and  the tomographic state reconstruction on Bob's.
Our results may open a way to new protocols for one-sided device-independent quantum key distribution~\cite{PRA85-010301} based on the DLCZ protocol employing single-rail qubits~\cite{Nature414-413}.
\\

\section{Methods}

{\small

\textbf{Experiment}
Here we present the experimental details of the scheme depicted in Fig.~\ref{fig-overall_schematic}. The source laser is a continuous-wave Ti:sapphire laser of 860 nm. The heralded single photons are conditionally produced based on the setup presented in Ref.~\cite{PRA87-043803} at an average count rate of about 8000 $\mathrm{s}^{-1}$ using a weakly pumped non-degenerate optical parametric oscillator (OPO). The single photons are impinged on a beam splitter characterized  by reflectivity $R$, which was set to four different values $R \in \{0.08, \, 0.38, \, 0.50, 0.90\}$. 

Alice and Bob perform homodyne measuments on the transmitted and reflected signals, respectively. A piezo electric transducer is used to control the relative optical path length of the LO of about 10 mW and the signal field, which determines the relative phase. 
This relative phase is locked, using an Integral (I) feedback control scheme, to one of the six values $\theta \in \{0, \pm \pi /6, \pm \pi /3, \pi /2\}$ for Alice, and is scanned across the full range $\phi \in [0, 2 \pi)$ for Bob. 
200,000 quadratures, each with an integration time of 500 ns, were recorded for each phase $\theta$, resulting in a total of 1,200,000 quadratures for each data set; the entire experiment took about 20 minutes for each data set. Four data sets were measured for each beam splitter reflectivity $R$.

\textbf{Single Photon Characterization}
The experimentally generated heralded single-photon input state to the beam splitter is well modeled by an incoherent mixture of vacuum, single-photon and two-photon contributions given by
\begin{equation}
\hat{\rho}_{in} = p_0 \ket{0} \bra{0} + p_1 \ket{1} \bra{1} + p_2 \ket{2} \bra{2}.
\label{initial_state_before_BS}
\end{equation}
The input state parameters are
$p_0 = 0.120 \pm 0.007$, $p_1 = 0.857 \pm 0.008$, and $p_2 = 0.02 \pm 0.01$
as measured by setting the beam splitter reflectivity to $R=1$, that is, sending all heralded single photons to Bob, and performing single mode tomography.
The higher order contributions add up to a negligible value of $p_{h} = 0.004 \pm 0.002$.

\textbf{Imperfections of Alice's Measurements}
The imperfections of Alice's measurements include the inefficiency and inaccuracy of homodyne detection as well as other losses in Alice's apparatus.
The inefficiency of Alice's homodyne detection of $1 - \eta_h = 0.04 \pm 0.01$ can be attributed to the imperfect spatial mode-match of $0.017 \pm 0.004$, homodyne detector circuit noise of $0.011 \pm 0.002$, and the inefficiency of photo-diodes (HAMAMASTU, S3759SPL) of $0.01$.
The inaccuracy of Alice's homodyne detection is mostly caused by phase fluctuations when locking the relative phase between the signal beam to be measured and the LO:
the root-mean-square phase-lock fluctuations are $\Delta \theta = 3.9^\circ$, estimated from the signal to noise ratio of the error signal feedback to the piezo electric transducer used to lock the relative phase to zero.
Other losses in Alice's apparatus are the propagation inefficiency from the OPO to the homodyne detector of $0.015 \pm 0.003$ and the losses induced by the imperfections of the anti-reflection coating in the beam splitter which are $1.07 \times 10^{-2} \pm 0.01 \times 10^{-2} $ at worst; these add up to an additional loss of $l_{\mathrm{A}} = 0.025 \pm 0.007$.

\begin{acknowledgments} 
This work was partly supported by the SCOPE program of the MIC of Japan, PDIS, GIA, G-COE, and APSA commissioned by the MEXT of Japan, FIRST initiated by the CSTP of Japan, ASCR-JSPS, and the Australian Research Council Centre of Excellence CE110001027. M.F. and S.T. acknowledges financial support from ALPS. M.Z. would like to acknowledge financial support from the European Union Seventh Framework Programme (FP7/2007-2013) under grant agreement n$^{\circ}$ [316244].
\end{acknowledgments}

\noindent
 \\ {\bf Author Contributions} \quad
H.M.W. and A.F. supervised the project.
H.M.W and M.Z. developed the theory and defined scientific goals.
M.F. designed, performed the experiment, acquired and analyzed data with the help of S.T.
M.F., H.M.W. and M.Z. wrote the manuscript with assistance from all other co-authors.

\section{SUPPLEMENTARY MATERIAL}

Here we assess the influence of the experimental imperfections on the violation of the EPR-steering inequality presented in the main text. We consider the experimental imperfections associated with the single-photon source, and the inefficiency and inaccuracy of Alice's homodyne detection.

\begin{figure}[!b]
\begin{center}
\includegraphics[bb=120 35 520 444, width=6cm]{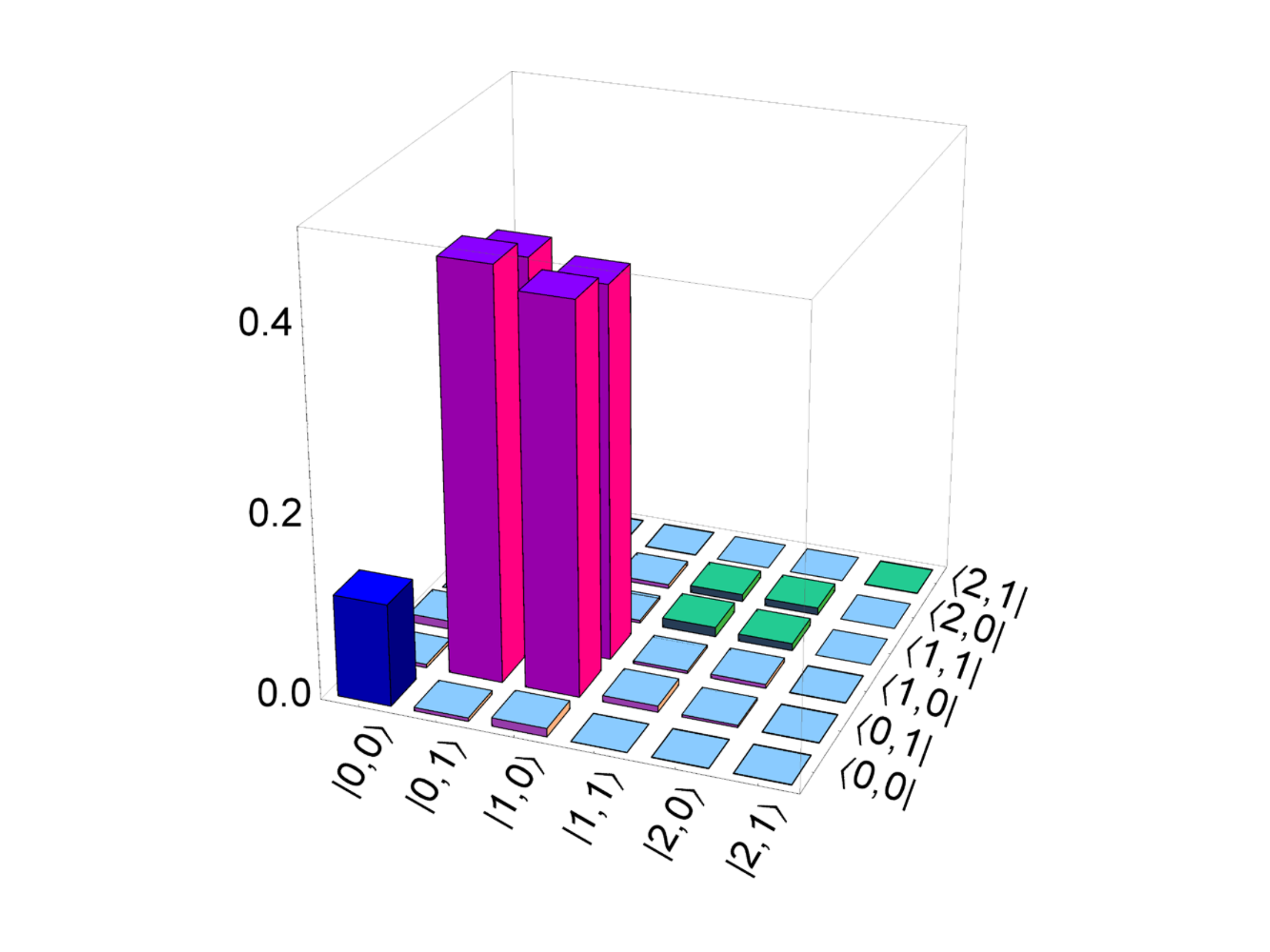}
\caption{
A typical two-mode tomographic reconstruction of the absolute values of the density matrix elements associated with the quantum state given in Eq.~(\ref{afterBS}) with $R=0.50$, truncated to the qubit subspace on Bob's side. The blue, magenta, and turquoise boxes show the density matrix elements of the vacuum, the single-photon, the two-photon and the higher-order subspaces, respectively.}
\vspace{-4mm}
\label{fig:qubit_qutrit_abs_R=50}
\end{center}
\end{figure}

{\bf State preparation}
Heraled single photons (modeled by Eq. 6) are impinged on the beam splitter, generating a mixed quantum state shared by Alice and Bob, which is given by
\begin{eqnarray}\label{afterBS}
\hat{\rho}_{\mathrm{AB}} &=& \uplus \sqrt{p_0} |00\rangle \notag \\
&&\uplus \sqrt{p_1} (\sqrt{R} |01\rangle - \sqrt{1-R} |10\rangle) \nonumber \\
&&\uplus \sqrt{p_2} (R |02\rangle - \sqrt{2 R (1-R)} |11\rangle + (1-R) |20\rangle)\, ,
\end{eqnarray}
where $R$ denotes the beam splitter reflectivity and $\uplus$ is defined as $\uplus \alpha |a \rangle \equiv +|\alpha|^2 |a \rangle\langle a|$~\cite{Jones06}. A typical two-mode tomographic reconstruction of this state with $R = 0.50$ is presented in Fig.~\ref{fig:qubit_qutrit_abs_R=50}. In the figure, the reconstructed state is truncated to the $\{\ket{0}, \ket{1}\}$ subspace on Bob's side because, as stated in the main text, we consider in this work a nonlinear EPR-steering inequality for \textit{qubits}.

{\bf Homodyne detection}
The local oscillators of Alice and Bob are synchronized by sourcing them from the same laser, as a continuous process during the experiment. This phase locking is necessary for our experiment to succeed, because if Alice's LO is not well phase-locked to Bob's, she will not be able to obtain good correlations, and therefore will not be able to demonstrate EPR-steering. Note that since this phase locking can in principle be done using only a classical channel, prior to the sharing of the single photon~\cite{Wiseman04laser}, it does not add any additional assumptions to our experiment.  

In our EPR-steering test, Alice performs homodyne measurements on her part of the mixed state $\hat{\rho}_{\mathrm{AB}}$ using inefficient photoreceivers. The inefficiency of her photoreceivers is modeled by introducing a photon loss channel, which is then followed by a fully efficient homodyne measurement. The quantum state shared by Alice and Bob subjected to the photon loss process on Alice's side can be written as
\begin{equation}
\hat{\rho}^{\mathrm{loss}}_{\mathrm{AB}} = \sum_{n=0}^{2} \hat{K}_{n} \hat{\rho}_{\mathrm{AB}} \hat{K}^{\dagger}_{n},
\end{equation}
where $\hat{K}_{n}$'s are three Kraus operators that describe the photon loss channel for $\hat{\rho}_{\mathrm{AB}}$. These operators are given by
\begin{eqnarray}
\hat{K}_{0} &=& |0 \rangle_{\mathrm{A}}\langle 0| + \sqrt{\eta_{\mathrm{A}}} |1 \rangle_{\mathrm{A}}\langle 1| + \eta_{\mathrm{A}} |2 \rangle_{\mathrm{A}}\langle 2|, \\
\hat{K}_{1} &=& \sqrt{1 - \eta_{\mathrm{A}}} |0 \rangle_{\mathrm{A}}\langle 1| + \sqrt{2 \eta_{\mathrm{A}}(1 - \eta_{\mathrm{A}})} |1 \rangle_{\mathrm{A}}\langle 2|, \\
\hat{K}_{2} &=& (1 - \eta_{\mathrm{A}}) |0 \rangle_{\mathrm{A}}\langle 2|
\end{eqnarray}
and correspond to three distinct possibilities of not losing any photons, losing a single photon and losing two photons, respectively. Here $\eta_{\mathrm{A}}=\eta_h(1-l_{\mathrm{A}})$, where $\eta_h = 0.96 \pm 0.01$ is the efficiency of Alice's photoreceivers and $l_{\mathrm{A}} = 0.025 \pm 0.007$ denotes other losses in Alice's apparatus. 
The theoretical fully efficient homodyne measurement, which follows the photon loss process is modelled by projecting $\hat{\rho}^{\mathrm{loss}}_{\mathrm{AB}}$ onto the quadrature eigenstate $|x^{\theta}_{\mathrm{A}}\rangle$.

We also allow for the fact that Alice's homodyning is not perfectly accurate by including in our analysis  the long-time phase-lock fluctuations, modeled by a Gaussian distribution centered around the phase lock point $\theta$. This gives us the following expression for Bob's (normalized) local quantum state conditioned on Alice's local oscillator phase $\theta$ and a coarse-grained result $s$
\begin{eqnarray}\label{conditionedstate}
\hat{\rho}^{\theta''}_{s} &=& \int_{-\infty}^{\infty} \frac{d\tilde{\theta}}{\sqrt{2 \pi (\Delta \theta)^2}} \exp\left[-\frac{(\tilde{\theta}-\theta)^2}{2 (\Delta \theta)^2}\right] \nonumber \\
&\times& \int_{\mathbb{R}_{s}} d x^{\tilde{\theta}}_{\mathrm{A}} \langle x^{\tilde{\theta}}_{\mathrm{A}}| \hat{\rho}^{\mathrm{loss}}_{\mathrm{AB}} |x^{\tilde{\theta}}_{\mathrm{A}}\rangle/P(s|\theta).
\end{eqnarray}
Here $\Delta \theta = 3.9^\circ$ is the measured value for Alice's root-mean-square phase-lock fluctuations and $P(s|\theta) = 0.5$.
Note that we had to integrate the coefficients of the conditioned state with respect to $x^{\tilde{\theta}}_{\mathrm{A}}$ over the negative ($\mathbb{R}_{-}$) and positive ($\mathbb{R}_{+}$) real line in order to coarse-grain on the sign $s$ of the quadrature result $x^{\tilde{\theta}}_{\mathrm{A}}$.

{\bf Bob's conditioned state}
Finally, we restrict Bob's local conditioned quantum state to the qubit subspace: $\hat{\rho}^{\theta}_{s} = \hat{\rho}^{\theta'}_{s}/\mathcal{N}$, where $\hat{\rho}^{\theta'}_{s}$ is the (unnormalized) restricted conditioned state obtained by discarding from $\hat{\rho}^{\theta''}_{s}$ all terms that contain two-photon contributions and
$\mathcal{N} = 
1 - p_{2} R^{2}$ is the renormalization factor.
We can explicitly write Bob's local conditioned quantum state as
\begin{widetext}
\begin{eqnarray}\label{restrictedconditionedstate}
\hat{\rho}^{\theta}_{s} &=& \{[p_{1} R + 2 p_{2} R(1-R)] \ket{1}\bra{1} + [p_{0} + p_{1} (1-R) + p_{2} (1-R)^{2}] \ket{0}\bra{0} \nonumber \\
&-& s \sqrt{\eta_{\mathrm{A}} R (1-R) 2/\pi} [p_{1} + p_{2} (1-R) (2-\eta_{\mathrm{A}})] e^{-(\Delta \theta)^{2}/2} (e^{i \theta} \ket{1}\bra{0} + e^{-i \theta} \ket{0}\bra{1})\}/(1 - p_{2} R^{2}). \nonumber \\
\end{eqnarray}
\end{widetext}

\begin{figure}[!b]	
\begin{center}
\includegraphics[bb=246 114 595 319, width=8.5cm]{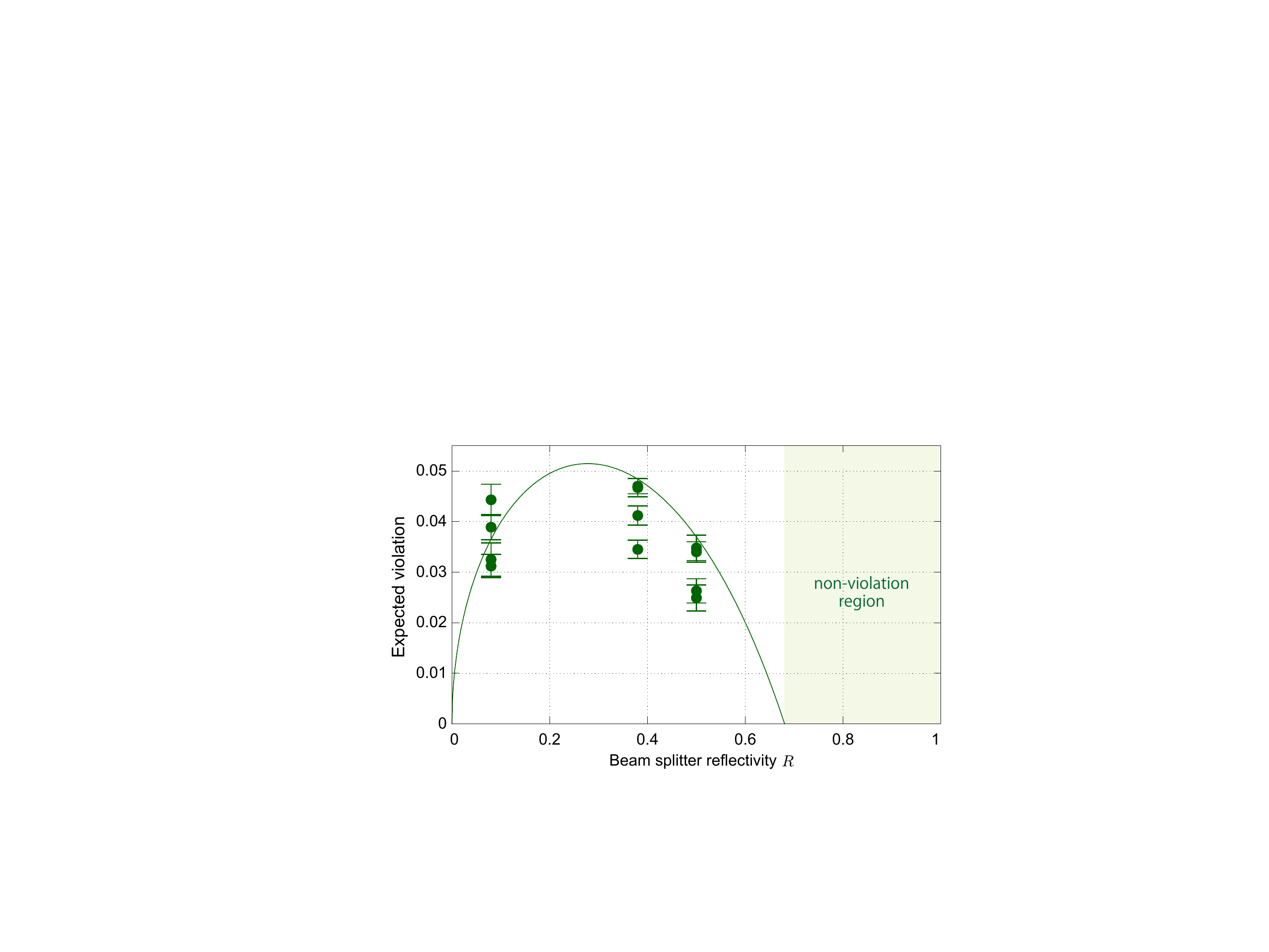}
\end{center}
\vspace{-4mm}
\caption{The expected violation of the EPR-steering inequality plotted as a function of the beam splitter reflectivity $R$. The solid line shows the theoretically predicted violation, which takes into the account the experimental imperfections. Four sets of data were taken for each $R$ to confirm reproducibility of the experimental results. To assess the tomographic errors, error bars were calculated using the bootstrap method. The discrepancies in the data sets can be attributed to the inconsistencies in the setup preparation, most significantly, the varying single-photon generation probability $p_{1}$. The observed violation is slightly smaller than the theoretically predicted value due to the imperfections in the phase estimation procedure during Bob's tomography.} \label{fig:inequality_violation}
\end{figure}

{\bf Expected violation of the EPR-steering inequality}
Given this state we can evaluate the expected value for the left-hand side of the EPR-steering inequality (5) defined in the main text:
\begin{eqnarray}
&&\frac{1}{n} \sum_{{j}=1}^{n} \sum_s P(s|\theta_{j})  \, s \, {\rm Tr}[\hat{\sigma}^{\theta_{j}}_{\mathrm{B}} \rho_s^{\theta_{j}}] \nonumber \\
&& = 2 e^{-(\Delta \theta)^{2}/2} \sqrt{\eta_{\mathrm{A}} R (1-R) 2/\pi} \ \frac{p_{1} + p_{2} (1-R) (2-\eta_{\mathrm{A}})}{1 - p_{2} R^{2}}. \nonumber \\
\end{eqnarray}
For the right-hand side of the EPR-steering inequality we also restrict Bob's unconditioned state to the qubit subspace:
\begin{eqnarray}\label{restrictedunconditionedstate}
\hat{\rho} &=& \{[p_{1} R + 2 p_{2} R(1-R)] \ket{1}\bra{1} \nonumber \\
&+& [p_{0} + p_{1} (1-R) + p_{2} (1-R)^{2}] \ket{0}\bra{0}\}/(1 - p_{2} R^{2}). \nonumber \\
\end{eqnarray}
The right-hand side of the EPR-steering inequality (5) thus evaluates to
\begin{eqnarray}
&& f(n) \sqrt{1 - \mathrm{Tr}\left[ \hat{\sigma}_{\mathrm{B}}^{\mathrm{z}} \hat{\rho} \right]^{2}} \nonumber \\
&& = f(n) \sqrt{ 1 - \frac{[p_{0} + p_{1} (1-2R) + p_{2} (1-R)(1-3R)]^{2}}{(1 - p_{2} R^{2})^{2}}}. \nonumber \\
\end{eqnarray}

We plot the expected violation of the EPR-steering inequality in Fig.~\ref{fig:inequality_violation}.
This manifests an asymmetry in the inequality violation in terms of the beam splitter reflectivity $R$.
If there were no imperfections in the state preparation then the degree of violation would actually be independent of $R$.
The chief imperfection causing this asymmetry is the less than unit efficiency $p_1 = 0.857 \pm 0.008 < 1$ of single-photon generation, as the value of the left-hand side decreases linearly with $p_1$, at least for small $p_2$, while the right-hand side takes its maximum value when $ p_1 R = 0.5 $. Consequently, there exists a maximum reflectivity $R_{\mathrm{max}} \approx 0.68$ for which the inequality is violated when $0 < R \leq R_{\mathrm{max}}$.

There is no minimum reflectivity for which the inequality is violated, however, there exists an optimum reflectivity $R_{\mathrm{opt}} < 0.50$ (in our experiment this optimal value is $R_{\mathrm{opt}} \approx 0.28$) for which the degree of violation is greatest.
This can be understood as follows. Alice is the party that is supposed to prove that she can steer Bob's local quantum state. Therefore, it is the efficiency and accuracy of Alice's homodyne measurements that matters the most for a successful demonstration of EPR steering. As a result, in an EPR-steering experiment we have to compensate the experimental imperfections on Alice's side by increasing the portion of the single photon that is send to her.


\end{document}